\documentclass[letters,usenatbib]{mnras}

\usepackage[T1]{fontenc}
\usepackage{ae,aecompl}

\usepackage{graphicx}	% Including figure files
\usepackage{amsmath}	% Advanced maths commands
\usepackage{amssymb}	% Extra maths symbols
\title[Theoretical study of an LAE-CIV absorption pair at $z =$ 5.7]{Theoretical study of an LAE-CIV absorption pair at $z =$ 5.7}

\author[L. A. Garc\'ia et al.]{
L. A. Garc\'ia$^{1,2}$\thanks{E-mail: lgarcia@swin.edu.au}, 
E. Tescari$^{2,3}$, E. V. Ryan-Weber$^{1,2}$ and J. S. B. Wyithe\,$^{2,3}$\\
\\
$^{1}$Centre for Astrophysics and Supercomputing, Swinburne University of Technology, Hawthorn, VIC 3122, Australia\\
$^{2}$ARC Centre of Excellence for All-Sky Astrophysics (CAASTRO)\\
$^{3}$School of Physics, The University of Melbourne, Parkville, VIC 3010, Australia
}

\date{Accepted XXX. Received YYY; in original form ZZZ}

\pubyear{2017}

\begin{document}
\label{firstpage}
\pagerange{\pageref{firstpage}--\pageref{lastpage}}
\maketitle

% Abstract of the paper
\begin{abstract}
We present a theoretical model to predict the properties of an observed $z =$ 5.72 Lyman $\alpha$ emitter galaxy -- \ion{C}{IV} absorption pair separated by 1384 comoving kpc/$h$. We use the separation of the pair and an outflow velocity/time travelling argument to demonstrate that the observed galaxy cannot be the source of metals for the  \ion{C}{IV} absorber. We find a plausible explanation for the metal enrichment in the context of our simulations: a dwarf galaxy with $M_{\star} =$ 1.87 $\times$ 10$^{9} M_{\sun}$ located 119 comoving kpc/$h$ away with a wind velocity of $\sim$ 100 km/s launched at $z \sim$ 7. Such a dwarf ($M_{\text{UV}} =$ - 20.5) is fainter than the detection limit of the observed example. In a general analysis of galaxy -- \ion{C}{IV} absorbers, we find galaxies with -20.5 $< M_{\text{UV}} <$ - 18.8 are responsible for the observed metal signatures. In addition, we find no correlation between the mass of the closest galaxy to the absorber and the distance between them, but a weak anti-correlation between the strength of the absorption and the separation of galaxy -- absorber pairs.
\end{abstract}


%%%%%%%%%%%%%%%%% BODY OF PAPER %%%%%%%%%%%%%%%%%%

\section{Introduction}

There is an intrinsic connection between the formation and evolution of galaxies in the Universe and the properties of the gas in the intergalactic medium (IGM), such as its metallicity and ionization state. These properties are successfully reproduced by simulations that employ feedback models, in forms of radiative and kinetic winds, able to chemically enrich the outer regions of galaxies \citep{shen2013}. The detection of galaxies that shape observations of the IGM provides information on the interplay of the outflowing winds from the star--forming regions and the gas in the outskirts that is photoionized and polluted by these feedback mechanisms.\newline
%The physical processes that take place during the galaxy evolution cause a dramatic change in the ionization state of the surrounding gas, but also, galaxies seem to be the main drivers of the chemical enrichment via feedback winds \citep{shen2013}. The detection of galaxies with associated observables laying in the IGM provides information of the interplay of the outflow winds from the star--forming regions and the gas in the outskirts that is photoionized and cooled by these feedback mechanisms.\newline
One of the best ways to explore the properties of the gas in the IGM is by the detection of \ion{C}{IV} absorption systems. Among other high--ionization states, \ion{C}{IV} traces low--density regions in the IGM at temperatures of $T \geqslant$ 10$^{4}$ K \citep{oppenheimer2009,tescari2011,cen2011,finlator2015,rahmati2016}. In addition, \ion{C}{IV} is a good indicator of ionized regions due to its large ionization potential energy and its detection in the spectra of background high redshift quasars \citep[e.g.][]{danforth2008,ryanweber2009, cooksey2010, simcoe2011,dodorico2013,boksenberg2015,dodorico2016} from $z \sim$ 0 to 6.\newline
At low and intermediate redshift, extensive observational campaigns are carried out with the goal to find, identify and characterize galaxy -- \ion{C}{IV} absorbers pairs:  at $z \sim$ 0.05 by \citet{burchett2016} and at $z \sim$ 2--3 \citep{steidel2010,turner2014}. Observing these galaxy -- \ion{C}{IV} absorber systems when the Universe is approaching the tail of Reionization ($z \sim$ 6),  might give us a hint of the connection between the galaxies that drove the Reionization and the metal enrichment in the very early Universe. For example, \citet{oppenheimer2009} explored the physical environment  of the absorbers assuming different scenarios for the ionizing background and found that \ion{C}{IV} absorbing gas is primarily intergalactic at $z \sim$ 6. The absorbers are distributed at distances up to 200 physical kpc from the parent galaxy. In their {\small{HM2001}} model, the \ion{C}{IV} absorbers are only associated with galaxies of $M_{\star} \sim$ 10$^{9}M_{\sun}$. In addition, work from \citet{oppenheimer2009,finlator2016} showed that local sources can affect the strength of \ion{C}{IV} absorption. \newline
With the detection of a high redshift ($z =$ 5.72) Lyman $\alpha$ emitting galaxy by \citet[][hereafter D15]{diaz2015} located 212.8 physical kpc/$h$ from a high column density (log N$_{\ion{C}{IV}}$(cm$^{-2}$)  $=$ 14.52) \ion{C}{IV} absorber \citep{dodorico2013}, there is now observational evidence of a galaxy -- absorber pair at this redshift. This detection opens the question: is the \ion{C}{IV} absorber due to an outflowing wind from the nearby star forming galaxy or does it instead arise due to an undetected dwarf galaxy closer to the line of sight?\newline
A theoretical comparison with different feedback prescriptions at $z =$ 6 by \citet{keating2016} finds that their models are unable to reproduce the configuration observed by D15, and that their strong absorbers are linked to nearby galaxies ($<$ 100 pkpc) with very high halo masses (log $M_{h}/M_{\sun} \geqslant$ 10). The low incidence rate of strong \ion{C}{IV} absorption systems makes it doubly difficult to simulate the observed scenario.\newline
This letter is devoted to exploring the likelihood of reproducing a system as detected by D15 and to giving a plausible explanation for the enrichment. In addition, we make some theoretical predictions in order to guide the analysis of future observations.\newline 
The work follows the methodology and models described in \citet[][MNRAS submitted; hereafter G17]{garcia2017}. The numerical simulations were run using a customized version of the smoothed particle hydrodynamics (SPH)  code {\small{GADGET-3}} \citep{springel2005}  with cosmological pa\-ra\-me\-ters from \citet{planck2015}: $\Omega_{0m}=$ 0.307, $\Omega_{0b}=$ 0.049, $\Omega_{\Lambda}=$ 0.693 and $H_0=$ 67.74 km s$^{-1}$Mpc$^{-1}$ (or $h =$ 0.6774). The simulated run Ch 18 512 MDW has comoving box size and softening of 18 Mpc/$h$ and 1.5 kpc/$h$, respectively, and includes 2$\times$512$^{3}$ dark matter and gas particles. We assume a momentum--driven wind feedback model with a wind mass--loading factor $\eta=2 \times \frac{\text{600 km s}^{-1}}{v_w}$ and a velocity of $v_w=2\sqrt{\frac{G M_{h}}{R_{200}}}= 2\times v_{\text{circ}}$.\newline
The simulation has been post--processed with a field radiation due to the cosmic microwave background (CMB) and the \citet{haardt2012} ultraviolet/X-ray background from quasars and galaxies with saw-tooth a\-tte\-nua\-tion (HM12), an effective prescription for HI self-shielding according to \citet{rahmati2013a}, a photoionization modeling for \ion{C}{IV} using {\small{Cloudy}} v8.1 \citep{ferland2013} for optically thin gas and Voigt profile fitting with the code {\small{Vpfit}} v.10.2 \citep{vpfit}. As described in G17, our numerical simulations correctly reproduce observed global statistics of \ion{C}{IV} absorbers, namely, the column density distribution function and the cosmological mass density $\Omega_{\ion{C}{IV}}$. Throughout the paper, we use the prefix c for comoving and p for physical distances. Please note that in our simulation the closest snapshot to the $z =$ 5.72 D15 system is at $z =$ 5.6.

\section{The LAE -- CIV absorption pair of D\'iaz et al. 2015}\label{lae_abs_pair}
In this section we investigate the likelihood of reproducing the galaxy -- \ion{C}{IV} absorber pair detected by D15 with theoretical models. Specifically, a CIV absorption system in the line of sight to the background quasar J1030+0524, with a column density of log N$_{CIV}$ (cm$^{-2}$) $=$ 14.52$\pm$0.08 \citep{dodorico2013} at a distance of 212.8 $^{+14}_{-0.4}h^{-1}$ pkpc (1384 ckpc/$h$ in the adopted cosmology) from a Lyman $\alpha$ emitter galaxy (LAE) of $M_{\text{UV}} =$ - 20.7 at $z =$ 5.72, with inferred parameters log ($M_{\star}/M_{\odot}$) $=$ 9.4 \citep[using the stellar mass function for $z =$ 6 galaxies from][]{song2016} and log ($M_{h}/M_{\odot}$) $ =$ 10 - 11 \citep[derived from $z \sim$ 6.6 LAE clustering  from][]{ouchi2010}. The possible scenarios that the authors propose to explain the metal enrichment of the region (in particular with Carbon) are either a very powerful galactic outflow that left the LAE at earlier times or an undetected dwarf galaxy closer to the absorber.\newline
In the simulations, the chemical enrichment is driven by supernovae that produce and expel metals to the outer regions of galaxies. Wind velocities in the adopted feedback model depend on the halo mass of the galaxies. Therefore, only the most massive objects produce winds powerful enough to pollute regions at more than 1 cMpc/$h$, but these objects are quite rare at high redshift, especially in small simulations. We calculate the wind velocity and travel time of the most massive galaxy in the box in Table~\ref{wind} to set an upper limit on the time required for an outflow to reach a region at a distance of 1300 ckpc/$h$ (comparable to the D15 example).
\begin{table}
\caption{Wind velocity ($v_w$) and travel time ($t$) to enrich a region at 1300 ckpc/$h$ from our most massive galaxy ($M_{h}$).}
\label{wind}
\centering
\begin{tabular}{cccc} 
\hline
Box size &  $M_{h}$ & $v_w$  & $t$ \\ 
(cMpc/$h$) &  ($\times$ 10$^{11} M_{\sun}$) & (km/s) &  (Gyr)\\ \hline
18 & 4.9  & 446  & 0.73\\ 
\hline
\end{tabular}
\end{table}
The results from Table~\ref{wind} suggest that not even the most massive galaxy is able to produce winds with the velocity required to travel 1300 ckpc/$h$ in a reasonable time. The time displayed in Table~\ref{wind} is comparable with the age of the Universe at $z =$ 5.6. Thus, for this option to be viable, galaxies would need to be formed at an age ($z \geqslant$ 30) incompatible with the current paradigm of galaxy formation. Therefore, we rule out  the possibility that an outflow produced by the LAE enriched a region at 1384 ckpc/$h$ in the context of the feedback model implemented in our simulations. Nevertheless, alternative configurations could lead to a different result: more aggressive prescriptions for supernova--driven outflows or larger halo masses that produce higher wind velocities. The latter condition could be fulfilled in our simulations with a galaxy halo mass of $\sim$ 1.5 $\times$ 10$^{12} M_{\sun}$. However, galaxies have only been observed with stellar masses up to 10$^{10} M_{\sun}$ at $z \sim$ 6 \citep{song2016}. Assuming a mass--to--light ratio of 10 (inferred from the simulations at this redshift), this scenario is still out of reach by at least an order of magnitude.\newline
This test also reveals interesting details of the model: the chemical enrichment is driven by galaxies nearby the absorbers, since galactic outflows cannot travel a distance larger than about 1 cMpc in less than 0.5 Gyr and chemically enrich the IGM. At this redshift, active galactic nuclei are the only sources energetic enough to produce winds that travel that far.\newline

\begin{figure}
\centering
\includegraphics[width=\columnwidth]{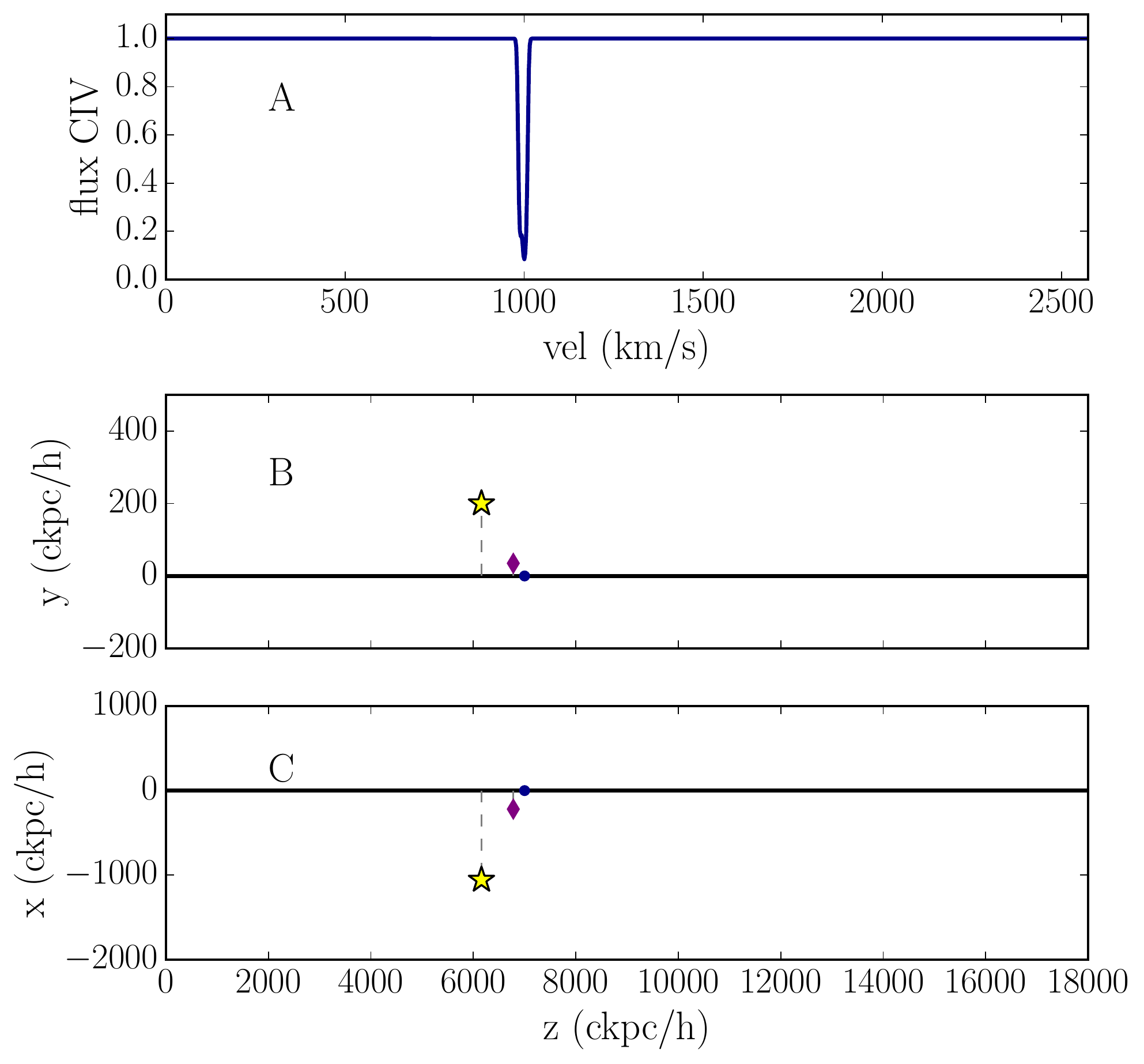}
\includegraphics[scale=0.4]{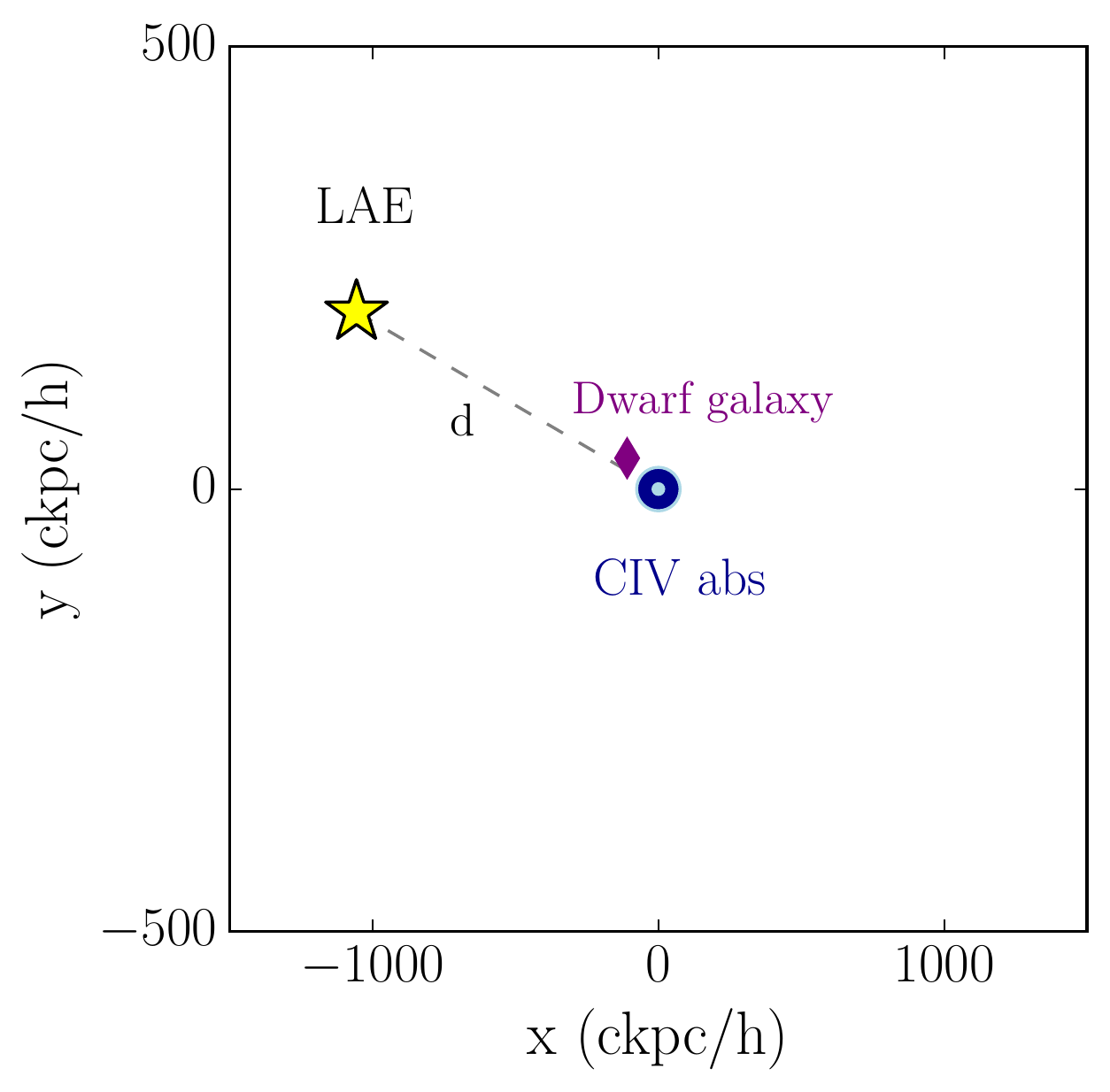}
\caption{Theoretical example of the observed LAE -- CIV absorption pair. In this window, we test one of the hypotheses proposed in D15: the existence of an undetected dwarf galaxy in the field of the LAE. We identify all the CIV absorptions at $z =$ 5.6 with log N$_{CIV}$ (cm$^{-2}$) in the range of 14--15 and the closest galaxies in the surroundings of each absorption. The upper panel (A) shows a feature with log N$_{CIV}$ (cm$^{-2}$) $=$ 14.3 $\pm$ 1.1 on a line of sight traced along the $z$ direction. Panel B shows the position of the strong CIV absorption (blue point), an LAE (yellow star) with $M_{h} =$ 1.54 $\times$ 10$^{10} M_{\sun}$ and a 3D distance of 1296 ckpc/$h$ from the absorption, and the closest galaxy in the field (purple diamond), a dwarf galaxy of $M_{\star} =$1.87 $\times$ 10$^{9} M_{\sun}$ at a 3D distance from the absorber of 119 ckpc/$h$, that could not have been detected in the observations. Panel C shows the same configuration in a different edge-on projection (xz). The bottom panel displays the physical disposition of the system face-on: the line of sight from the quasar (in the background) contains the absorption and it is orthogonal to the plane of the image.}
\label{fig:cand_dwarf}
\end{figure}

Next, we explore the dwarf galaxy scenario, in which the presence of another galaxy closer to the absorber, and fainter than the detection limits for optical observations, is responsible for the enrichment. The upper limit for an undetected galaxy is: $M_{\text{UV}} =$ -20.5, corresponding to $M_{\star} = $ 10$^{9.3} M_{\sun}$ \citep{song2016}. Independent narrow band observations of Ly$\alpha$ emission set a limit on the star formation rate, SFR $=$ 5-10 $M_{\sun}$/yr. The goal of this analysis is to find mock scenarios that resemble the conditions of the LAE detected in D15 and, if possible, confirm or rule out the hypothesis of an undetected dwarf galaxy.\newline
At $z =$ 5.6, the simulated galaxies have halo masses in the range of 7.83 $<$ log $M_h (M_{\sun}) <$ 10.83. We classify them in the following categories: LAE candidates ($M_{h} \geq$ 1.48 $\times$ 10$^{10} M_{\sun}$) and  dwarf galaxies (1.48 $\times$ 10$^{9} M_{\sun}$ $< M_{h} < $ 1.48 $\times$ 10$^{10} M_{\sun}$) to provide a fair comparison with the observational parameters. Galaxies with halo mass below 1.48 $\times$ 10$^{9} M_{\sun}$ are excluded from the statistics because of their low resolution. We stress that we do not follow the Ly$\alpha$ emission of any galaxy. The criterion to describe the galaxies is based only on their stellar and halo masses.\newline
Stochastically, 1000 galaxies are selected from the friends--of--friends catalogue and from each one, a line of sight is projected with a random impact parameter $d$ up to 1500 ckpc/$h$. The  column density N$_{CIV}$ and the position of the absorption in the box is recovered in each case. We focus on CIV absorption systems with log N$_{CIV}$(cm$^{-2}$) in the range of 14-15. We then look for all the galaxies around the absorption (our absorption spectra have an uncertainty of 7 km/s, or $\pm$ 49 ckpc/$h$, along the line of sight). This set of galaxies is identified and sorted by mass and distance (the 3D distance can be decomposed in components parallel and perpendicular to the line of sight). The array in mass allows us to tag the galaxy as LAE or dwarf (or just exclude it according to the resolution criterion mentioned above), whereas the distance array defines the closest galaxies to the absorption. The cross--matching of the arrays leaves 4 systems with a configuration LAE -- CIV absorption (-- dwarf galaxy), close to the observational arrangement. \newline
In particular, one system drew our attention for its geometry being closest to the D15 observations. We show the configuration LAE -- CIV absorption -- dwarf galaxy in Figure~\ref{fig:cand_dwarf}. The CIV absorption feature has log N$_{CIV}$ (cm$^{-2}$) $=$ 14.3 $\pm$ 1.1, the LAE has $M_{h} =$ 1.54 $\times$ 10$^{10} M_{\sun}$ at a 3D distance of 1296 ckpc/$h$ from the CIV system. The closest galaxy in the field, a dwarf galaxy of $M_{\star} =$ 1.87 $\times$ 10$^{9} M_{\sun}$, $M_{h} =$ 9.67 $\times$ 10$^{9} M_{\sun}$, SFR = 0.07 $M_{\sun}$/yr is located at a 3D distance from the absorber of 119 ckpc/$h$. A dwarf galaxy with this stellar mass lies just below the 50\% complete value for $M_{\text{UV}}$ images of D15 at $M_{\star} =$ 2 $\times$ 10$^{9} M_{\sun}$. In this case, the velocity of an outflow produced by the dwarf galaxy is $v_w \sim$ 100 km/s, leading to a travel time of $\sim$ 260 Myr (\textit{i.e.} the outflow was launched at $z \sim$ 7).\newline
The spatial configuration displayed in Figure~\ref{fig:cand_dwarf} favors an undetected dwarf galaxy in the observations of D15 and confirms that, in the context of our feedback model, the chemical enrichment is caused by nearby galaxies. In addition, these results emphasize that the metal absorption line systems offer the best technique for detecting galaxies beyond the limits of imaging at high redshifts. \newline

\begin{figure}
\centering
\includegraphics[width=\columnwidth]{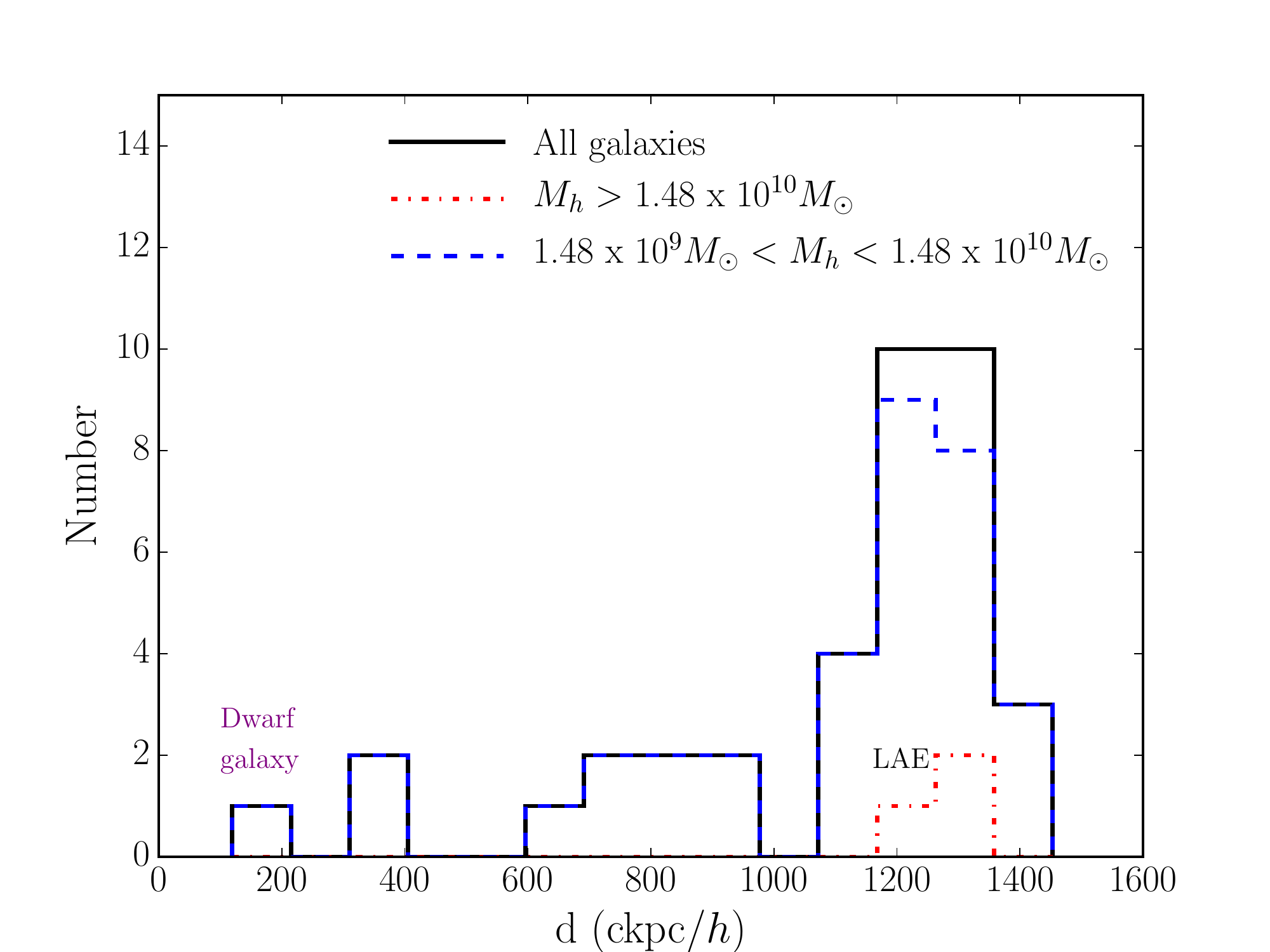}
\caption{Distribution of galaxies around the CIV strong absorption shown in Figure~\ref{fig:cand_dwarf} up to a 3D distance of 1500 ckpc/$h$. The black histogram accounts for all galaxies, whereas the red dashed--dotted line shows the systems with masses $M_{h} \geq$ 1.48 $\times$ 10$^{10} M_{\sun} $ (LAE) and the blue dashed line the contribution from dwarf galaxies 1.48 $\times$ 10$^{9}$ $< M_{h} / M_{\sun} < $ 1.48 $\times$ 10$^{10}$ in the box. The galaxies below the mass resolution limit of 1.48 $\times$ 10$^{9} M_{\sun}$ are not shown.}
\label{fig:histo_230}
\end{figure}

%Furthermore, the short amount of cosmic time between the first episodes of star formation and $z \sim$ 6 means that diffuse gas in the IGM, and especially the CGM, is probably polluted by single galaxies.
%Nonetheless, configurations as shown in Figure~\ref{fig:cand_dwarf} are extremely rare in a sample of a 1000 sightlines, thus metals yields must come from different sources and mix, to produce the metallicity measured in the simulations.
As a final point, we mention that in the D15 observations, there is a small probability that an associated galaxy of any mass is obscured by the quasar. The highest resolution image of this quasar is an HST/ACS z-band image taken by \citet{stiavelli2005}. Conservatively, we estimate that the quasar light would prevent the detection of galaxies within a radius of 1 arcsec. Thus, 11\% of the circular area within 18 pkpc/$h$ (119 ckpc) from the quasar line of sight is obscured. At this redshift, a separation of 18 pkpc/$h$ corresponds to 3.07 arcsec. 
In order to understand the incidence rate of galaxies around the CIV absorption in the particular scenario discussed above, we plot the distribution of 3D distances from the absorption to each galaxy in a range of 1500 ckpc/$h$ in Figure~\ref{fig:histo_230}. There are many more dwarf galaxies closer to the CIV absorber than the LAE.
Furthermore, the probability of finding a randomly placed dwarf galaxy at 119 ckpc/$h$ from a metal absorber is 5.1\%, knowing that the number density of dwarf galaxies in the sample is 0.46 objects per cMpc$^{3}$ and the volume where these pairs are likely to occur is about 0.02 cMpc$^{3}$.

\section{Galaxy -- absorber connection}
\begin{figure}
\centering
\includegraphics[width=\columnwidth]{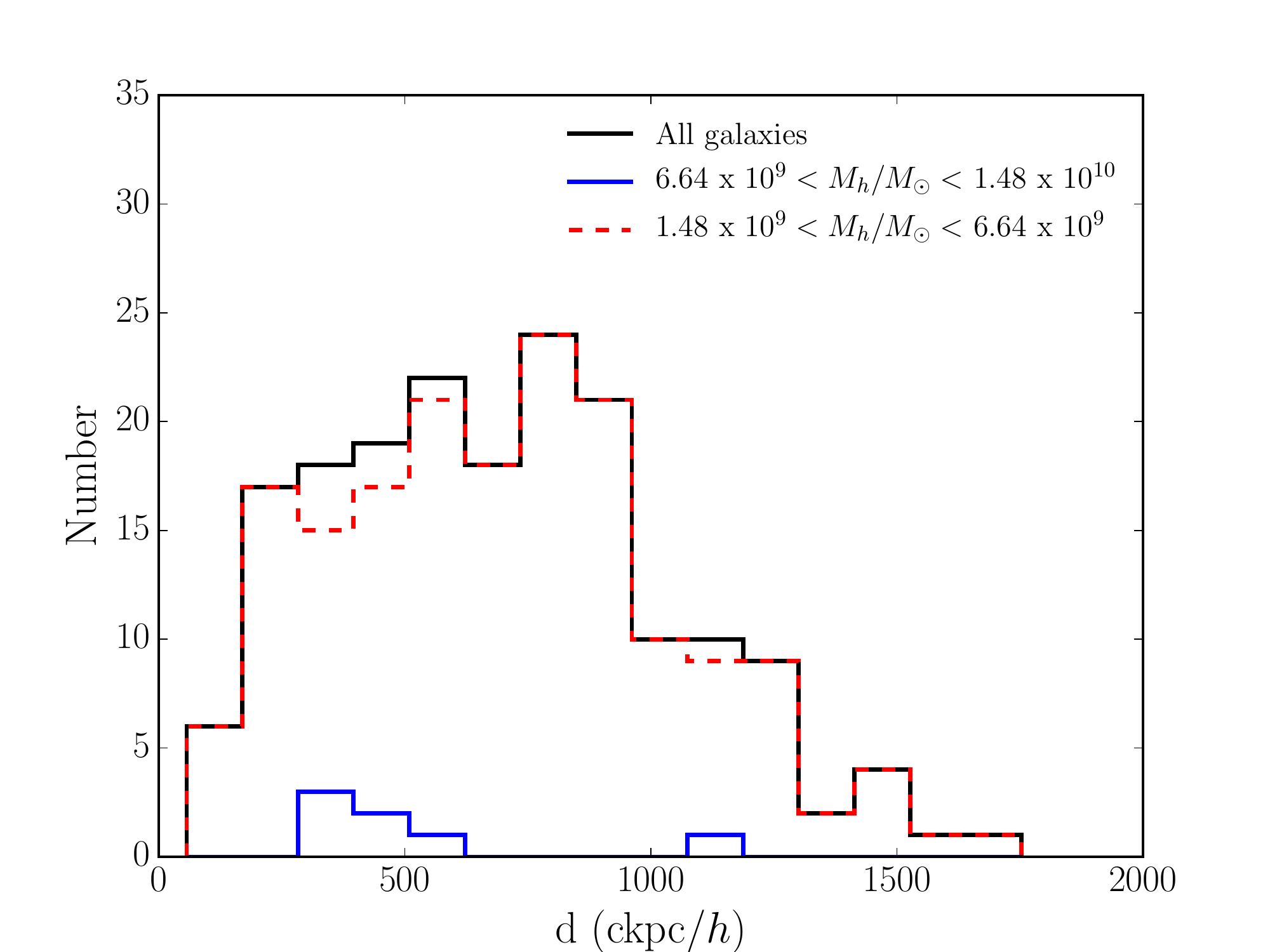}
\caption{Mass distribution of the 3D distance from the CIV absorptions to the closest galaxy. The black histogram accounts for all the galaxies. There are not high mass galaxies ($M_{h} \geq$ 1.48 $\times$ 10$^{10} M_{\sun}$), meaning that, in all cases, the closest galaxies from the absorptions lay in the category of dwarfs. Intermediate mass galaxies are presented in blue solid line (7) and low mass galaxies with dashed red (175). The galaxies below the mass resolution limit 1.48 $\times$ 10$^{9} M_{\sun}$ are not presented. In cases where $d \geqslant$ 1500 ckpc/$h$, the enrichment is most likely due to an unresolved galaxy.}
\label{fig:histo_mass}
\end{figure}

\begin{figure}
\centering
\includegraphics[width=\columnwidth]{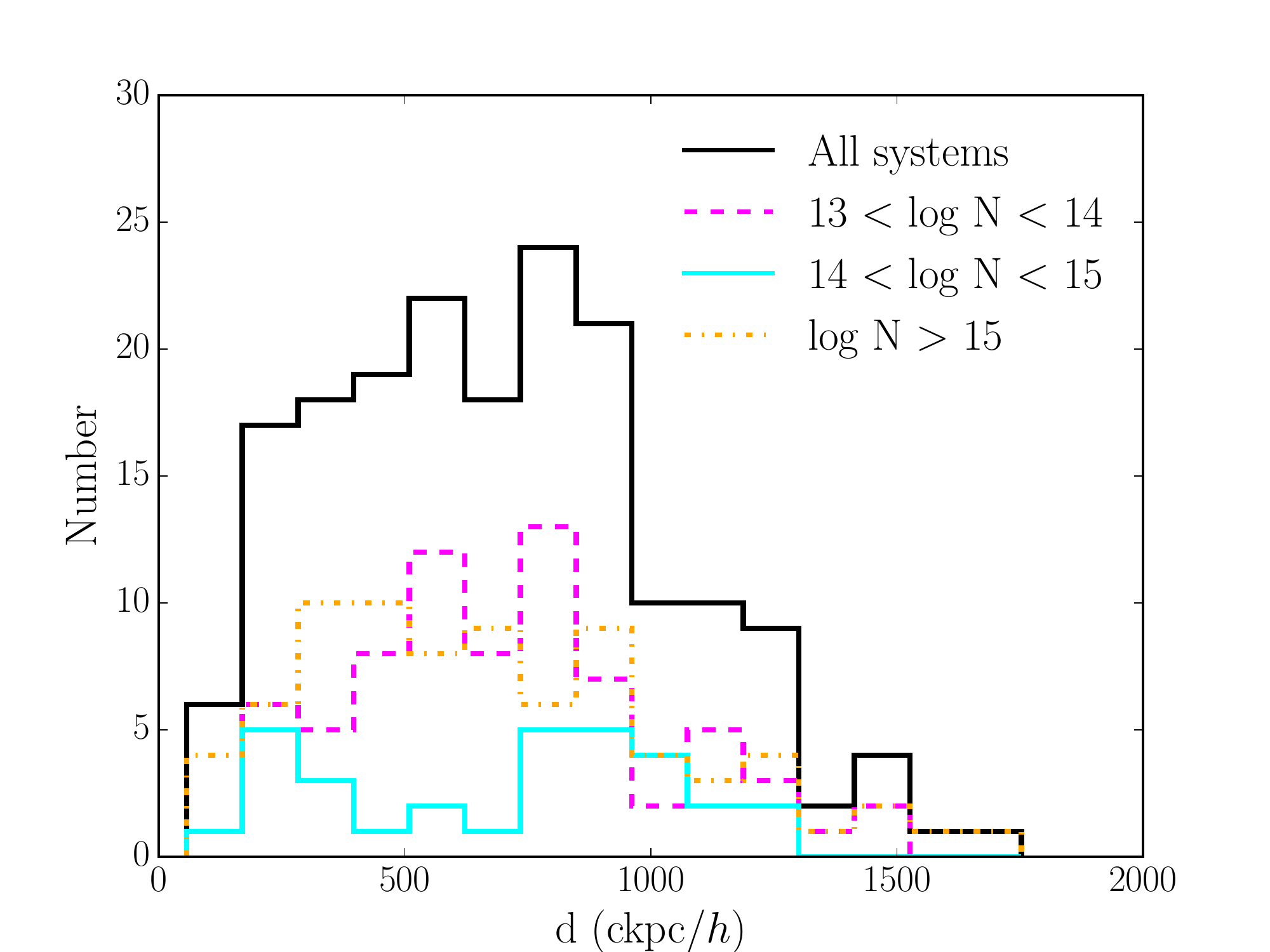}
\caption{Distribution of the 3D distance from the CIV absorptions to the closest galaxy with respect to the column density of the absorption. The black histogram accounts for all the column densities in the range of 13 $\leq$ log N (cm$^{-2}$) $\leq$ 16. The magenta dashed line shows systems with 13 $\leq$ log N (cm$^{-2}$) $<$ 14, the cyan solid line the systems with 14 $\leq$ log N (cm$^{-2}$) $<$ 15 and in orange dashed--dotted line systems with column densities log N (cm$^{-2}$) $\geq$ 15.}
\label{fig:histo_N}
\end{figure}

We have shown a theoretical realization of the observed LAE -- CIV absorption pair at $z =$ 5.6. Now, we want to explore the physical connection of these systems (if any), using the properties of the galaxies and the column densities of the absorbers in the simulated box. For the purpose of the analysis of this section, which is  made with galaxies in the simulations above the mass resolution limit, we define high ($M_{h} \geq$ 1.48 $\times$ 10$^{10} M_{\sun}$), intermediate (6.64 $\times$ 10$^{9}$ $< M_{h} / M_{\sun} < $ 1.48 $\times$ 10$^{10}$) and low mass (1.48 $\times$ 10$^{9}$ $< M_{h} / M_{\sun} < $ 6.64 $\times$ 10$^{9}$) galaxies.\newline
The connection of the galaxy -- CIV systems can be studied using galaxy -- absorber pairs with log N$_{CIV}$ (cm$^{-2}$) $>$ 13 and taking into account the nearest galaxy to the absorption. We implicitly assumed that this nearest galaxy is the main source of metal enrichment for the absorber. \newline
We construct the overall distribution of the distance from each absorption to its closest galaxy in the simulated box. From the initial sample of 1000 lines of sight, 182 CIV absorbers have a galaxy companion above the mass resolution. In Figure~\ref{fig:histo_mass}, we display this distribution with respect to the mass of the galaxy. \newline
There are not high mass galaxies in the random selection from the friends--of--friends catalogue because just 2\% of the objects are in the range of mass $M_{h} \geq$ 1.48 $\times$ 10$^{10} M_{\sun}$, which makes it extremely hard to encounter a random line of sight through these galaxies. On the other hand, almost 90\% of the galaxies in the box are intermediate and low mass (1.48 $\times$ 10$^{9}$ $< M_{h} / M_{\sun} < $ 1.48 $\times$ 10$^{10}$). Therefore, it is not surprising that the closest galaxy to each N $>$ 10$^{13}$ cm$^{-2}$ CIV absorber has a mass in this range. \newline
%Therefore, it is not surprising that galaxies in the high mass category ($M_{h} \geq$ 1.48 $\times$ 10$^{10} M_{\sun}$) have \textit{0 abs/28 los used/235 in the initial sample of halos}, whereas ``high--mass'' dwarfs with 6.64 $\times$ 10$^{9}$ $< M_{h} / M_{\sun} < $ 1.48 $\times$ 10$^{10}$ have associated \textit{7 abs/116 los used/262 in the initial sample of halos} and ``low--mass'' dwarf galaxies with 1.48 $\times$ 10$^{9}$ $< M_{h} / M_{\sun} < $ 6.64 $\times$ 10$^{9}$ produced \textit{175 abs/826 los used/8345 in the initial sample of halos}. 
Additionally, intermediate and low mass galaxies are typically found at a mean distance of 520 ckpc/$h$ (78.8 pkpc/$h$) and 700 ckpc/$h$ (106.1 pkpc/$h$), respectively. This result is only partially in agreement with findings of \cite{oppenheimer2009}, for high stellar--mass galaxies in their {\small{HM2001}} model: strong CIV absorptions at $z =$ 6.0 are associated with galaxies with $M_{\star} \sim$ 10$^{9} M_{\sun}$ and the typical galaxy -- absorption separation is less than 100 pkpc. It is important to clarify that, in addition to the intrinsic difference between {\small{HM2001}} and HM12 backgrounds, their highest mass galaxies are close to our mass resolution (lower) limit. Nonetheless, we confirm that CIV is mostly found in the IGM \citep[G17;][]{oppenheimer2009}. 
Finally, Figure~\ref{fig:histo_mass} reveals that the CIV strong absorptions at high redshift are driven by dwarf galaxies with $M_{\star}$ spanning the range 10$^{8.46-9.42} M_{\sun}$ and SFR$=$0.01--2.5 $M_{\sun}/$yr. These parameters allow us to infer an absolute magnitude $M_{\text{UV}}$ in the range of (-20.5,-18.8) \citep[using the $M_{\star}-M_{\text{UV}}$ relation at $z \sim$ 6 from][]{song2016} that should be achieved with future surveys. The detection and study of these faint galaxies is fundamental, since they are believed to provide the largest contribution of photons to complete the Reionization of Hydrogen \citep{robertson2015, lui2016}. \newline
%\textit{Also, the metal and radiative feedback are linked to the same stars that produce metals and emit the most ionizing photons.} 
Figure~\ref{fig:histo_N} shows the distribution of 3D distance from the CIV absorptions to the closest galaxy as a function of column density of the absorption. The 182 absorbers span the range of 13 $\leq$ log N (cm$^{-2}$) $\leq$ 16. From this sample, there are 73 absorbers with 13 $\leq$ log N (cm$^{-2}$) $<$ 14, 31 in the range of 14 $\leq$ log N (cm$^{-2}$) $<$ 15 and 78 with log N (cm$^{-2}$) $\geq$ 15.
The mean distance in the three cases is respectively: 710.6, 705.5 and 680.6 ckpc/$h$. This result indicates that there is a weak negative correlation between the column density of the CIV absorptions and the separation of absorber -- galaxy pairs.\newline 
Finally, it is worth noting that strong CIV absorptions in our simulations are produced not only by very close structures with large masses ($M_{h} \sim$ 10$^{10-11} M_{\odot}$, which we rarely find within this box size), but also for small galaxies. 

\section{Conclusions} \label{conclusions}
We have explored the likelihood of reproducing the observed LAE -- CIV absorption pair detected by D15 and studied the physical processes that produced the metal enrichment in the IGM at $z \sim$ 5.6. In the context of our feedback model, we rule out the scenario of an outflow produced by the observed LAE at a distance of 1384 ckpc/$h$ at early times.\newline 
%and impose constraints on the age and mass of LAE that would be likely to cause the enrichment of the region where the absorption is observed via feedback winds.\newline
Instead, our simulations support a scenario in which a dwarf galaxy, undetected in the field, very close to the absorber (119 ckpc/$h$), with $M_{\star} =$ 1.87 $\times$ 10$^{9} M_{\sun}$, $M_{h} =$ 9.67 $\times$ 10$^{9} M_{\sun}$ and SFR = 0.07 $M_{\sun}$/yr, is responsible for the CIV absorption.\newline
Additionally, we find that the main drivers of the chemical enrichment of the IGM and CIV strong absorptions at $z =$ 5.6 are dwarf galaxies, which are mostly observed at a mean distance of 700 ckpc/$h$ (106.1 pkpc/$h$). These galaxies have stellar masses $M_{\star}$ in the range 10$^{8.46-9.42} M_{\sun}$ and SFR$=$0.01-2.5 $M_{\sun}/$yr. From these parameters, we infer an absolute magnitude $M_{\text{UV}}$ in the range of (-20.5,-18.8), derived by combining results of \citet{song2016} with the Kennicut relation. Future observations and investigations of these faint galaxies are fundamental, since they are believed to be the largest contributors of photons to the Reionization of Hydrogen.\newline
When exploring the galaxy -- absorber connection at high redshift, we find no correlation between the mass of the closest galaxy to the absorber and the distance between them. On the other hand, we encounter a weak negative correlation between the column density and the separation of the galaxy -- absorber pairs. The largest column densities are preferentially seen when separations are smallest. \newline

Parts of this research were conducted by the Australian Research Council Centre of Excellence for All-sky Astrophysics (CAASTRO), through project number CE110001020. This work was supported by the Flagship Allocation Scheme of the NCI National Facility at the ANU. The authors acknowledge CAASTRO for funding and allocating time for the project \textbf{Diagnosing Hydrogen Reionization with metal absorption line ratios} (fy6) during 2015 and 2016. We also thank the anonymous referee for their insightful comments. ERW acknowledges ARC DP 1095600.

\bibliographystyle{mnras}
\bibliography{thebib} 

%%%%%%%%%%%%%%%%%%%%%%%%%%%%%%%%%%%%%%%%%%%%%%%%%%
% Don't change these lines
\bsp	% typesetting comment
\label{lastpage}
\end{document}